\documentclass[%
 reprint,
 amsmath,amssymb,
 aps,
American Journal of Physics,
]{revtex4-1}

\usepackage{graphicx}
\usepackage{dcolumn}
\usepackage{bm}

\begin{document}

\preprint{APS/123-QED}

\title{Some Static Properties of Slinky}

\author{A. Eskandari-asl}
\email[Email: ]{a{\_}eskandariasl@sbu.ac.ir}
\email{amir.eskandari.asl@gmail.com}
\affiliation{Department of physics, Shahid Beheshti University, G. C. Evin, Tehran 1983963113, Iran}

\date{\today}

\begin{abstract}
In this paper we use a simple discrete model for Slinky to explore some of its static properties. We derive some relations for vertically and U-shaped suspended Slinkies, based on which, some demonstrations are proposed that can be simply done in freshmen physics classes. 
\end{abstract}

\pacs{Valid PACS appear here}
\maketitle

\section{Introduction}
A flexible long spring called \textit{Slinky} is a very popular toy. There are many videos on the web, demonstrating interesting properties of falling Slinkies. In addition to its entertaining aspects, Slinky can be used for simple but interesting demonstrations for freshmen physics students, supported with simple theoretical analysis.

There are many papers investigating different properties of Slinky. In some of these papers, a suspended Slinky is studied \cite{french1994suspended,heard1977behavior,sawicki2002static,gluck2010project,essen2010static,mak1987static}, while in others dynamical properties such as wave propagation and free fall are considered \cite{gluck2010project,bowen1982slinky,calkin1993motion,young1993longitudinal,gardner2000slinky,
graham2001analysis,aguirregabiria2007falling,cross2012modeling}. In these studies, mostly a continuous model is considered, however, some authors use a discrete model \onlinecite{sawicki2002static,essen2010static}. For example, in Ref.\onlinecite{sawicki2002static} a simple discrete model is considered and some of the properties of a suspended Slinky are explored.

A more sophisticated discrete model can be found in Ref. \onlinecite{wilson2001energy}, where they consider instead of point-like masses, bars that are connected with springs. In Ref. \onlinecite{holmes2014equilibria} which is an extension of \onlinecite{wilson2001energy}, the bars are connected with three axial, rotational and shear springs.

In this paper we consider a simple discrete model for Slinky, like the one used in references \onlinecite{sawicki2002static} and \onlinecite{essen2010static}, and after restating some of their results for a vertically suspended Slinky, consider a U-shaped suspension. Based on our theoretical results, we propose some demonstrations that can be done in freshmen physics classes. In a real class at Shahid Beheshti Univ., we actually did these demonstrations, which seemed to be very attractive and inspiring for the students. 

The paper is organized as follow. In Sec.II, we consider vertically suspended Slinky and meanwhile introduce the discrete model. In Sec.III, U-shaped suspension from equal-height points is described, while Sec.IV considers suspension from two points with different heights. Finally, Sec. V concludes our work.   

\section{Vertical Suspension}
\begin{figure}[ht!]
\includegraphics[width=4.cm]{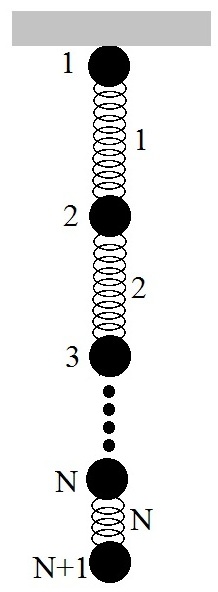}
\caption{\label{fig1} Discrete model for Slinky in vertical suspension.}
\end{figure}

In order to study vertically suspended Slinky, we model it with a collection of $ (N+1) $ small point-like objects with mass $ m $, connected with $ N $ ideal springs with spring constant $ k $, as shown in Fig.\ref{fig1}. At the limit of large $ N $, this model describes a spring with mass $ M_{s}=\left( N+1\right)  m $ and spring constant $ K_{s}=k/N $ .
When Slinky is suspended from one end, because of its mass, the upper parts are stretched more than the lower parts.  The change of length of the $ n $-th spring is shown by $ \delta_{n} $. Using the cancellation of the forces on each object, considering the appropriate boundary condition and doing some straight forward calculations, one can easily show that the length of the suspended spring is given by
\begin{eqnarray}
L=L_{0}+\sum_{n=1}^{N} {\delta_{n}}=L_{0}+\frac{M_{s} g}{2K_{s}} ,
\label{eq5} 
\end{eqnarray}
Where $ L_{0} $ is the equilibrium length of Slinky. One of properties that makes Slinky so special is that its equilibrium length is negligible, i.e., $ L_{0} \ll \frac{M_{s} g}{K_{s}} $. Therefore, length of a suspended spring is approximately $ L=\frac{M_{s} g}{2K_{s}} $. Based on this relation, we can do the following demonstration. First we suspend a Slinky from a fixed point and measure its length. Then, we use half of the Slinky and see that since the mass is halved and the spring constant is doubled, the length is decreased by a factor of four.

Now consider a downward vertical $ y $ axis with its origin at the suspension point. Position of the $ n $-th object is $ y_{n}=\frac{n}{N} L_{0}+\sum_{i=1}^{n-1} {\delta_{i}} $.For the limit of large $ N $, in every small element of length we would have many objects and springs. This means that the discrete model can be considered as a continuous one. For this continuous model, neglecting $ L_{0} $, one arrives at the following formula for the mass distribution over length
\begin{eqnarray}
\sigma \left( y \right) =\frac{K_{s}}{g} \sqrt{\frac{L}{L-y}}.
\label{eq7} 
\end{eqnarray} 
This equation satisfies the relation $ M_{s}=\int_{0}^{L} \sigma \left( y \right) dy $. Moreover, one can easily show that for negligible $ L_{0} $, the center of mass is located at 
\begin{eqnarray}
y_{cm}=\frac{1}{ M_{s}}\int_{0}^{L} y \sigma \left( y \right) dy=\frac{2L}{3}.
\label{eq8} 
\end{eqnarray} 
As a classroom demonstration, one can drop the Slinky at the same time with an object located at $ y_{cm} $. It is expected that the object and Slinky reach the floor at the same time. This can be seen more clearly by filming the experiment and watching the slow motion.

\section{U-Shaped Suspension from points with  equal heights}\label{s3}
\begin{figure}[ht!]
\includegraphics[width=8.cm]{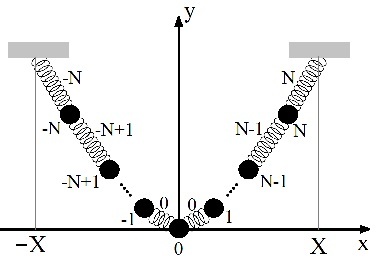}
\caption{\label{fig2} Discrete model for Slinky in U-shaped suspension with equal-height suspension points.}
\end{figure}

In this section, we consider Slinky suspended from both ends, with two suspension points at equal heights. For this case, our model is shown in Fig.\ref{fig2}. In this case, we have $ (2N+1) $ objects with mass $ m $ and $ 2 N+2 $ springs with spring constant $ k $, therefore, $ M_{s}= (2N+1) m $ and $ K_{s}=\frac{k}{2N+2} $. We label the objects and springs from $ -N $ to $ N $ and define the $ x-y $ coordinates as shown in Fig.\ref{fig2}. 

\begin{figure}[ht!]
\includegraphics[width=5.cm]{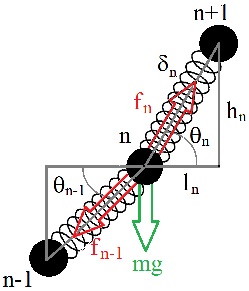}
\caption{\label{fig3} Forces exerted on the $ n $-th object.}
\end{figure}

\begin{figure}[ht!]
\includegraphics[width=5.cm]{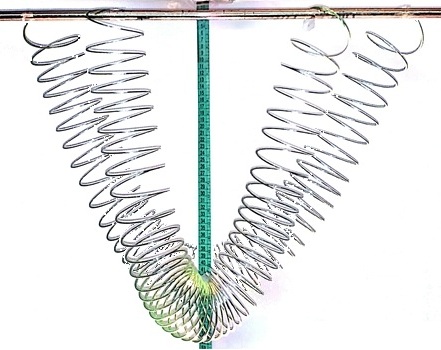}
\caption{\label{fig4} The depth of the U-shaped Slinky is independent of the distance between equal height suspension points.}
\end{figure}

In Fig.\ref{fig3} we depict the forces exerted on the $ n $-th object ($ 0<n<N $). Cancellation of the forces yields the following two relations
\begin{eqnarray}
&&f_{n} \sin{\theta_{n}}=m g+ f_{n-1} \sin{\theta_{n-1}}, 
\label{eq9} 
\end{eqnarray}
and
\begin{eqnarray}
&&f_{n} \cos{\theta_{n}}=f_{n-1} \cos{\theta_{n-1}}. 
\label{eq10} 
\end{eqnarray}
where $ f_{n} $ is the force exerted by the $ n $-th spring. We consider the case in which $ L_{0} $ is very small, that is, the equilibrium lengths of the springs in the model are approximately zero. As a consequence, $ \sin{\theta_{n}}=\frac{h_{n}}{\delta_{n}} $ and $ \cos{\theta_{n}}=\frac{l_{n}}{\delta_{n}} $. Combining these with the boundary condition $ 2 f_{0} \sin{\theta_{0}}=m g  $ (which is obtained for the object at origin), one can easily show that Eqs.\ref{eq9} and \ref{eq10} result in
\begin{eqnarray}
&&h_{n} =\frac{m g}{k} \left( n+\frac{1}{2}\right), 
\label{eq11} 
\end{eqnarray}  
and
\begin{eqnarray}
&&l_{n} =\frac{X}{N+1}, 
\label{eq12} 
\end{eqnarray}  
where $ X $ is half the distance between the suspension points, as indicated in Fig.\ref{fig2}. The coordinates of the $ n $-th object at the positive half are $ x_{n}=\sum_{i=1}^{n} l_{i} $ and $ y_{n}=\sum_{i=1}^{n} h_{i} $. Using Eqs. \ref{eq11} and \ref{eq12} and taking the limit of large $ N $, one can show that the shape of Slinky is a parabola with the following equation
\begin{eqnarray}
y(x) =\frac{M_{s} g}{8 K_{s} X^{2}} x^{2}. 
\label{eq13} 
\end{eqnarray} 
Lets define the depth of this parabola to be $ D=y(X)-y(0) $. From Eq.\ref{eq13} we see that $ D=\frac{M_{s} g}{8 K_{s} } $, and it is independent of $ X $. As a demonstration one can hold Slinky with their two hands and show that by changing the distance of the hands the depth of the formed parabola won't change (obviously, this distance between the suspension points can not get so huge that the Hook's law for the springs breaks down).This behavior is shown in Fig.\ref{fig4}. It is also noticeable that $ D $ is equal to the length of a halved Slinky which is vertically suspended (compare with Eq.\ref{eq5}).  

This result is not intuitive because of our daily experience with strings, for which the depth obviously depends on the distance between the suspension points. The difference can be understood by noting that the origin of this result in Slinky is the fact that we can ignore the equilibrium length of the small springs in our model, which means $ L_{0}\ll L $. However, the usual string lies on the other extreme, in which the spring stiffness is huge and $ L\approx L_{0} $.  

\section{U-Shaped Suspension from points with different heights}
\begin{figure}[ht!]
\includegraphics[width=8.cm]{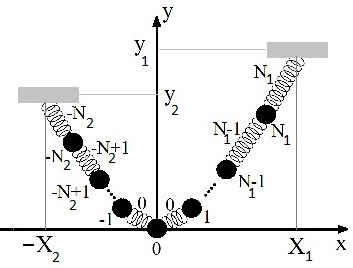}
\caption{\label{fig5} Discrete model for Slinky in U-shaped suspension with different-height suspension points.}
\end{figure}

The more general U-shaped Slinky in which the heights of the suspension points are different, is considered in this section. The corresponding discreet model is depicted in Fig.\ref{fig5}.
Following similar steps as the former case, one can obtain the $ y $ coordinate as a function of $ x $. Alternatively, we can use the results of Sec.\ref{s3} as follows. Eq.\ref{eq10} in combination with $ \cos{\theta_{n}}=\frac{l_{n}}{\delta_{n}} $ prove that the horizontal distance between the objects are equal. As a result, the part of Slinky which is at the right side of origin represents a spring with mass $ M_{1}=\frac{2 X_{1}}{X_{1}+X{2}} M_{s} $ and spring constant $ K_{1}=\frac{X_{1}+X{2}}{2 X_{1}} K_{s} $, where $ X_{1} $ and $ X_{2} $ are shown in Fig.\ref{fig5}. From Eq.\ref{eq13}, the $ y(x) $ equation for this part is given by
\begin{eqnarray}
y(x) =\frac{M_{1} g}{8 K_{1} X_{1}^{2}} x^{2}=\frac{M_{s} g}{2 K_{s} (X_{1}+X_{2})^{2}} x^{2}. 
\label{eq14}    
\end{eqnarray} 
The same reasoning applies to the part on the left side of origin and the same $ y(x) $ equation would be obtained. 

Eq.\ref{eq14} says that the functional form of $ y(x) $ just depends on the horizontal distance between the suspension points and not on the vertical distance. Therefore, identical Slinkies which are suspended from points with equal horizontal distances are all different parts of the same parabola. It should be mentioned that in our analysis we considered that the difference of the heights of the suspension points is not more than $ \frac{M_{s} g}{2 K_{s}} $, so that the valley lies between them, however, using the discrete model it can be easily shown that the result is not restricted to this condition.  

\section{Conclusions} 
In conclusion, we considered a simple discrete model for Slinky and applied it to cases of vertically and U-shaped suspended Slinkies.

In the first case, we obtained the length of Slinky and the location of its center of mass. We proposed two demonstrations for this part. In one of the demonstrations we halve the Slinky and show that its suspended length is divided by four. In the second demonstration we drop an object, simultaneous with Slinky, at the same height as its center of mass, and students can see that the center of mass motion is a simple free fall.  

In the case of U-shaped Slinky with equal-height suspension points, we obtained its shape and showed that it was a parabola. On the other hand, we showed that the depth of this parabola is independent of the horizontal distance of the suspension point (as long as the distance between the suspension points is not too much, so that the Hook's law holds for our spring). This non-intuitive result can be shown in a simple demonstration.

In a more general case, we considered a U-shaped Slinky with different-height suspension points and showed that the parabolic equation for its shape is just determined by the horizontal distance between the suspension points. Therefore, identical Slinkies which are suspended from points with equal horizontal distances are all different parts of the same parabola. 

Based on simple and beautiful physics contained in Slinky, we propose it to be used as a pedagogical device in freshmen basic physics.

\acknowledgments
I acknowledge A.R.Qalavand, the laboratory staff of Physics Department, Shahid Beheshti Univ., physics students who followed the discussions and helped providing some of the demonstrations, specially, Arya Gholampour, Amirmohammad Zakeri, Arash Daneshvarnejad, Farzad Nikzadian, Mohammadsharif Sadeghi, all of my Basic Physics students from the Faculty of Computer Science and Engineering, and also the Scientific Association of Physics Students, Shahid Beheshti Univ.



\bibliography{apssamp}

\end{document}